\documentclass[aps,prl,twocolumn,noshowpacs]{revtex4}
\usepackage{graphicx,epsfig, subfigure}
\usepackage{amsbsy}
\usepackage{amssymb}
\usepackage{amsmath}
\usepackage{xifthen} 
\graphicspath{{figures//},{//}}

\date{\today}

\begin{document}

\newcommand{\eqnref}[1]{Eq.~\ref{#1}}
\newcommand{\figref}[2][]{Fig.~\ref{#2}\ifthenelse{\isempty{#1}}{}{(#1)}}
\newcommand{\citeref}[1]{Ref. \cite{#1}}

\newcommand{\Sperp}{S}
\newcommand{\Q}{Q}

\medmuskip=2mu
\thinmuskip=2mu
\thickmuskip=2mu

\title{Parametric Excitation and Squeezing in a Many-Body Spin System}
\author{T.M. Hoang, M. Anquez, B.A. Robbins, X.Y. Yang, B.J. Land, C.D. Hamley,  and M.S. Chapman}

\affiliation{School of Physics, Georgia Institute of Technology,
  Atlanta, GA 30332-0430}

\begin{abstract}

We demonstrate a new method to coherently excite and control the quantum spin states of an atomic Bose gas using parametric excitation of the collective spin by time varying the relative strength of
the Zeeman and spin-dependent collisional interaction energies at multiples of the natural frequency of the system. Compared to the usual
single-particle quantum control techniques used to excite atomic spins (e.g. Rabi oscillations using rf or microwave fields), the method demonstrated here is intrinsically many-body, requiring inter-particle interactions. While parametric excitation of a classical system is ineffective from the ground state, we show that in our quantum system, parametric excitation from the quantum ground state leads to the generation of quantum squeezed states.
\end{abstract}

\maketitle

Ultracold atomic gases with well-characterized collisional interactions allow new explorations of non-equilibrium dynamics of quantum many-body physics and for synthesis of strongly-correlated quantum states including spin-squeezed  \cite{Gross10,Riedel10,Lucke11,HamleyNature2012} and  non-Gaussian entangled states \cite{GervingNature2012,Gring12,Strobel14} relevant for quantum sensing \cite{Ma11} and quantum information \cite{Braunstein05}. Here, we demonstrate a new method to coherently excite and control the quantum spin states of an atomic Bose gas using parametric excitation of the collective spin by time varying the relative strength of the Zeeman and spin-dependent collisional interaction energies at multiples of the natural frequency of the system. Compared to the usual single-particle quantum control techniques used to excite atomic spins (e.g. Rabi oscillations using rf or microwave fields), the method demonstrated here is intrinsically many-body, requiring inter-particle interactions. While parametric excitation of a classical system is ineffective from the ground state \cite{Mechanics}, we show that, in our quantum system, this leads to exponential excitation and the generation of quantum squeezed states.

\begin{figure}[t!]
	\begin{center}
	\begin{minipage}{3.5in}
		\includegraphics[scale=1]{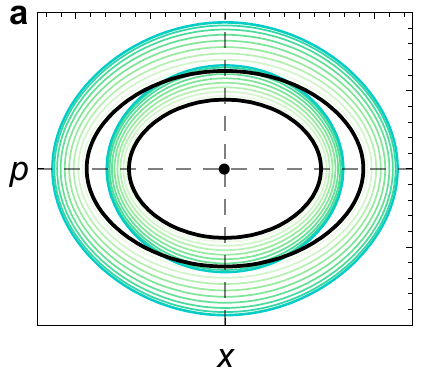}	
		\includegraphics[scale=1]{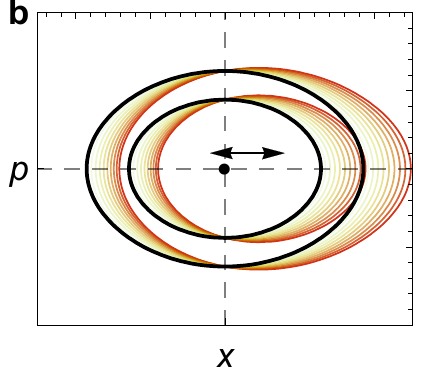}	\\		
		\includegraphics[scale=1]{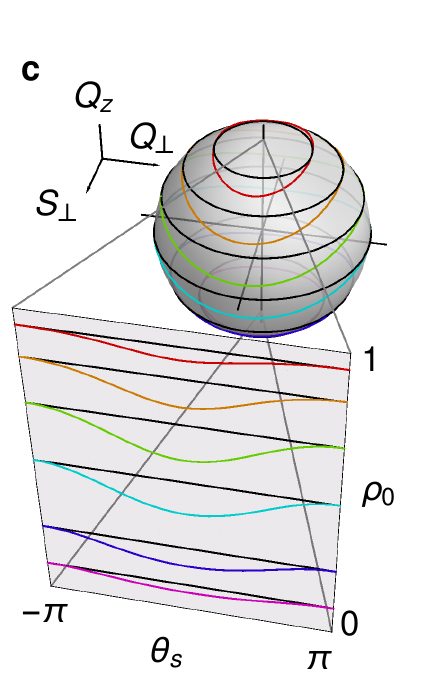}	
		 \includegraphics[scale=1]{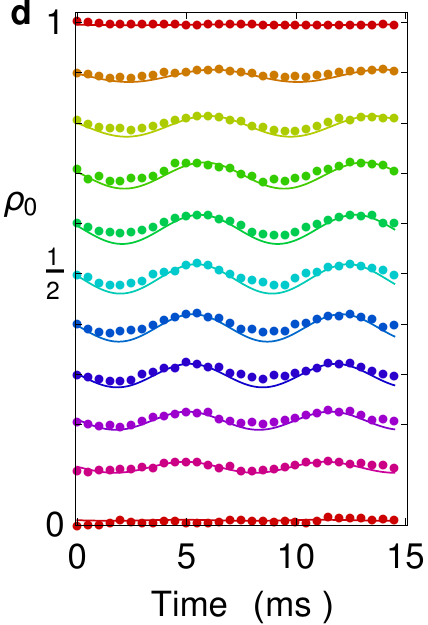}		 
	\end{minipage}
	\end{center}
    \caption{\textbf{Experimental concept.}  Momentum-position phase space of a harmonic oscillator under (a) parametric  and (b) direct excitation. The original orbits are shown in black, and the modified orbits are shown in color at different instants of the periodic excitation. (c) The phase space of the condensate. The collective states (normalized to $N$) lie on a unit sphere with axes ${S_\bot, Q_\bot, Q_z}$. The orbits of constant energy for non-interacting ($c=0$) spins are lines of latitude shown in black and the orbits for interacting spins with $q=10|c|$ are shown in color. The $\rho_0,\theta_s$ diagram is a Mercator projection of the hemisphere. $\theta_s=\theta_{+1}+\theta_{-1}-2\theta_{0}$ is the relative phase of the three Zeeman sub-levels, $m_F=0,\pm1$  (d) Measurements of the natural oscillations of $\rho_0$ at $B=1$~G for different initial states $\rho_0(t=0)\in[0,1]$. The experimental data (markers) are compared to simulations (lines). Unless otherwise indicated, the uncertainty in the measurement of $\rho_0$ is $<$1\%.}
\label{fig:Concept}
\end{figure}

Parametric excitation of an oscillating physical system can be achieved by periodically varying one of its parameters to modulate the natural frequency of the oscillator, $f_0$; a textbook example is a simple pendulum excited by modulating its length, $\ell$, such that $f_0(t) \propto 1/ \sqrt{\ell(t)}$ \cite{Mechanics}. A fundamental distinction between parametric excitation and direct excitation by periodic forcing is shown in Fig. \ref{fig:Concept}, which shows instantaneous phase space orbits of a simple oscillator for the two cases. For direct excitation, the applied force periodically displaces the equilibrium position of the oscillator, leaving the orbits otherwise unchanged, and efficient excitation occurs when the excitation frequency matches the natural frequency of the oscillator, $f=f_0$. For parametric excitation, the parameter modulation leaves the equilibrium location unchanged but instead periodically distorts the phase orbits;  in this case, efficient excitation occurs for excitation frequencies $f=2f_0/n$, $n=1,2,3...$ 

In ultracold atom traps, parametric excitation of the atomic motion, achieved by modulating the trapping potential, is used to measure the trap frequency as well as in a variety of studies including the excitation of Bose-Einstein condensate (BEC) collective density modes \cite{JinPRL1996,MewesPRL1996,EnglesPRL2007,JaskulaPRL2012}, controlling the superfluid/Mott insulator transition \cite{StoferlePRL2004,ZenesiniPRL2009}, and photon-assisted tunneling in modulated optical lattices and super-lattices \cite{SiasPRL2008,AlbertiNP2009,MaPRL2011,ChenPRL2011,HallerPRL2010,EckardtPRL2005}.

In this work, we demonstrate parametric excitation of the \emph{internal} states of a collection of atomic spins. The spins are coherently excited to non-equilibrium states  by a simple modulation of the magnetic field magnitude at very low frequencies ($<$200 Hz) compared to the energy difference of the Zeeman states ($\Delta E/h=0.7$ MHz, where $h$ is Planck's constant). The excitation spectrum is fully characterized and compares well to theoretical calculations. Parametric excitation of the ground state is also investigated. Classically, parametric modulation of an oscillating system does not excite the ground state.  Here, we show that the finite quantum fluctuations of the collective spin leads to parametric excitation of the ground state which manifest as exponential evolution of the fluctuations and the generation of non-classical squeezed states. The exponential evolution and squeezing of the spin fluctuations are measured and agree qualitatively with theory. Finally, we discuss how these techniques can be applied to related systems including the double-well Bose-Hubbard model and interacting (psuedo)spin-1/2 ensembles.  

The experiments use $^{87}$Rb Bose condensates with $N=40000$ atoms in the $F=1$ hyperfine level tightly confined in optical traps such that spin domain formation is energetically suppressed and dynamical evolution of the system occurs only in the internal spin variables. The Hamiltonian describing the  evolution of this collective spin system in a bias magnetic field $B$ along the $z$-axis is  
\cite{HoPRL1998,OhmiJPSJ1998,LawPRL1998,HamleyNature2012}:
\begin{equation}
\hat{H}=\tilde{c}\hat{S}^2-\frac{1}{2}q \hat{Q}_z
\label{Hamiltonian}
\end{equation}
where $\hat{S}^2$ is the total spin-1 operator and $\hat{Q}_z$ is proportional to the spin-1 quadrupole moment, $\hat{Q}_{zz}$. The coefficient $\tilde{c}$ is the collisional spin interaction energy per particle integrated over the condensate and $q=q_z B^2$  is the quadratic Zeeman energy per particle with $q_z=72$ Hz/$\mathrm{G}^2$ (hereafter, $h=1$). The  longitudinal magnetization $ \langle \hat{S}_z \rangle $ is a constant of the motion ($=0$ for these experiments); hence the first order linear Zeeman energy $p \hat{S}_z$ with $p \propto B $  can be ignored.
The spin-1 coherent states  can be represented on the surface of a unit sphere shown in \figref[c]{fig:Concept}with axes  $ \{\Sperp_\bot ,  \Q_{\bot } , Q_z \}$ where $\Sperp_\bot$ is the transverse spin, $S_\bot^2 = S_x^2+S_y^2 $, $Q_{\bot}$ is the transverse off-diagonal nematic moment,  $Q_\bot^2 =  Q_{xz}^2 + Q_{yz}^2$, and $Q_z = 2\rho_0-1$ where $\rho_0$ is the fractional population in the $F=1,m_F=0$ state. In this representation, the dynamical orbits are the constant energy contours of $ \mathcal{H} = \frac{1}{2}cS_\bot^2 - \frac{1}{2}q Q_z $ where $c=2N\tilde{c}$.

The experiment is conducted at high fields where the Zeeman energy dominates the spin interaction energy, $q/|c| = 10$. In this regime, the lowest energy state is the polar state ($\rho_0=1$) located at the top of the sphere, and the dynamical orbits of the excited states to leading order are simple rotations about the $Q_z$ axis with a frequency $ f_0 \approx q+c Q_z$  (Supplemental Information). Despite the small relative  magnitude of the spin interaction term, it has the important effect of breaking the polar symmetry and thereby slightly distorting the orbits from the latitudinal lines of the sphere. As the state orbits the sphere, the population $\rho_0$ undergoes small periodic nutations  at twice the orbit frequency, as shown in the $\rho_0,\theta_s$ projection in \figref[c]{fig:Concept}. The maximum nutation amplitude is $\Delta \rho_0 \approx 0.02$ for $\rho_0=0.5$ and goes to zero for $\rho_0=0,1$. Measurements of these distortions are shown in \figref[d]{fig:Concept} for different initial values of $\rho_0$.  

\begin{figure}[t!]
	\begin{center}
	\begin{minipage}{3.5in}
		\includegraphics[scale=1]{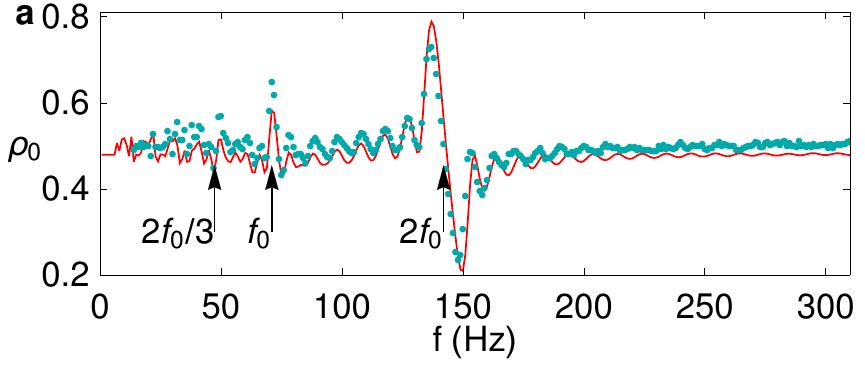}	
		\\
		\includegraphics[scale=1]{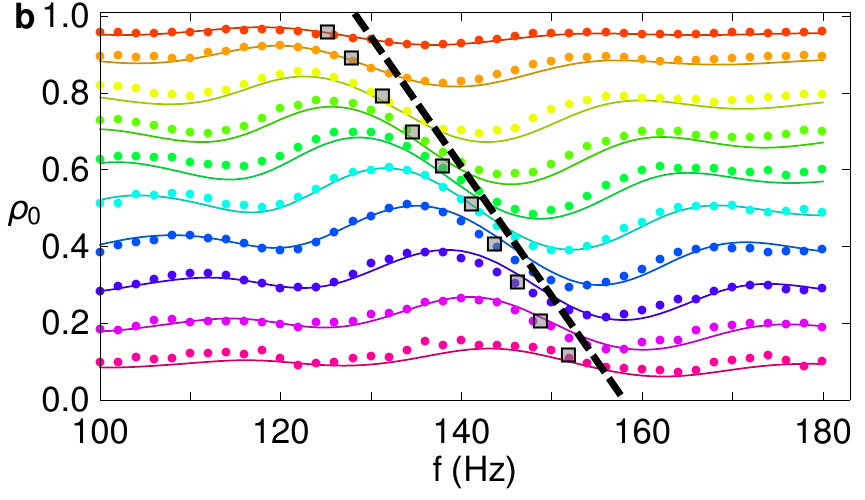}
		\includegraphics[scale=1]{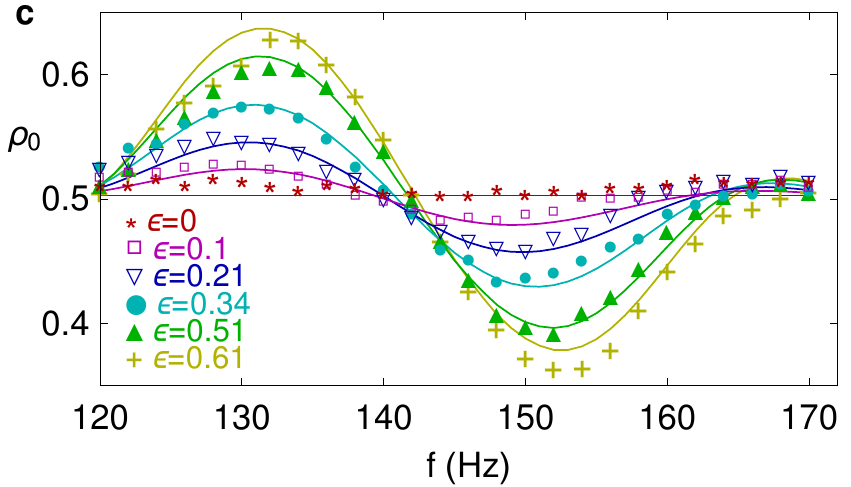}	
	\end{minipage}
	\end{center}
    \caption{\textbf{Demonstration of parametric excitation}. 
    (a) Population $\rho_0$ after 100~ms of parametric excitation for  $(\rho_0(0),\phi_0,\epsilon )=(0.5, \pi,\epsilon = 0.5)$. Data (markers) are plotted with simulation (solid line) for comparison. The excitation spectrum shows clear resonances at frequencies  $2f_0$ and $f_0$ 
(b) Populations $\rho_0$ after 40~ms of parametric excitation for different initial $\rho_0$ populations for $\phi_0=\pi$ and $\epsilon = 0.5$. Data (markers) are compared to simulation (solid lines), and hue colors correspond to the initial $\rho_0\in [0,1]$. The dashed line is the theoretical prediction for $2f_0$ resonance, and the square boxes indicate the location of the measured resonance. 
    (c) Population $\rho_0$ after 40 ms of parametric excitation for different modulation amplitudes $\epsilon$. 
    }
\label{fig:ParaExcitation}
\end{figure}

\noindent
\textbf{Parametric excitation}

Parametric excitation requires periodic modulation of one of the parameters of the Hamiltonian; in a spin-1 condensate described by Eq.~\ref{Hamiltonian}, this is  conveniently achieved by  modulating the bias magnetic field and hence the quadratic Zeeman energy term, $q(t) \propto B^2(t)$. The condensate is first prepared in a coherent state with $\rho_0, \theta_s=(0.5,\pi)$ at a field of 1~G, corresponding to an initial quadratic Zeeman energy $q_0=72~\mathrm{Hz}$. The spinor dynamical rate is $c=- 7(1)~\mathrm{Hz}$, determined from measurements of coherent oscillations  at low fields. 
To parametrically excite the spins, the magnetic field is modulated for a duration of time, after which the  spin populations are  measured to determine the final value of $\rho_0$.
The applied modulation is harmonic in $q$ and has the form  $q(t)=q_0 [1+ \epsilon \sin (2\pi f t-\phi_0)]$.
The  measured excitation spectrum versus modulation frequency is shown in \figref[a]{fig:ParaExcitation}. The spectrum shows the characteristic features of parametric excitation, namely strong excitation at $2f_0=142$~Hz and weaker excitation at $f_0$. Other resonances are theoretically observable at smaller $f=2f_0/n$ values; however, they are dominated by the tails of the more prominent peaks making them difficult to detect. The experimental data (marker) are compared to a simulation using Eq.~\ref{Hamiltonian} (solid line) and show good agreement overall.

Beyond comparing the experimental results to numerical solutions of the quantum Hamiltonian, insight into the parametric excitation is obtained by considering the mean-field  dynamical equations for $\rho_0$ and the quadrature angle $\theta = \theta_s/2$  \cite{ChangNature05}:
	\begin{eqnarray}
	\dot{\rho}_0 &=& 4 \pi c \rho_0(1-\rho_0)\sin 2\theta \notag \\
	\dot{\theta} &=& -2\pi[q - c(1-2\rho_0)(1+ \cos 2\theta)] \notag	
	\end{eqnarray}
These equations are similar to bosonic Josephson junction equations describing the double-well condensate \cite{RaghavanPRA1999,VardiPRL2001} and can be solved like-wise by integrating the phase and using the Jacobi-Anger expansion (Supplemental Information),
\begin{eqnarray}
	\label{Eqn:Rho0Freq}
	\dot{\rho}_0 &=& 4\pi c\rho_0(1-\rho_0) \sum_{n} J_n \left( \frac{4\pi q_m}{\omega} \right)   \sin[(n \omega-2 \omega_0)t+\phi_n] \notag
\end{eqnarray}
where $J_n$ is the Bessel function of order $n$, $q_m=\epsilon q_0$ is the modulation strength and $\phi_n$ depends on the initial conditions including the phase of the modulation (Supplemental Information). The parametric resonance frequencies are obvious from this solution because the time-average of $\dot{\rho}_0$ is zero unless   $\omega = 2 \omega_0 /n$. 

In the $q \gg |c|$ high field regime of these experiments, the collisional interactions shift the natural oscillation frequency  as $f_0=|\dot{\theta}|/2\pi \approx q+c(2\rho_0-1)$ to lowest order. This shift is investigated in \figref[b]{fig:ParaExcitation}, where the excitation is measured for different initial values of $\rho_0$. The hue colors correspond to the initial $\rho_0$ values using the same scale as the coherent oscillation data in \figref[d]{fig:Concept}, and the square markers  indicate the positions of the measured resonance frequencies where $\Delta\rho_0\approx0$.   
The measured resonance frequencies are in good overall agreement with the expected dependence on the initial value of $\rho_0$ shown with the dashed line---the small discrepancy is attributed to an inductive delay in the excitation of $\sim$1~ms (see Methods) that creates a phase offset $\delta \phi_0\sim 0.9$ rad in the excitation. Indeed, the experimental data compare very well to simulations (solid line) that include this phase offset.

The dependence of the excitation amplitude on the drive strength $q_m=\epsilon q_0$  is reflected in the Bessel function $J_n(4\pi q_m/\omega)$. In \figref[c]{fig:ParaExcitation}, the excitation is measured for different modulation amplitudes. The experimental data (markers) are compared to the simulations (solid line) and show good agreement (see Methods). As expected, the modulation amplitude does not affect the resonance frequency of the parametric excitation; however, increasing the  modulation amplitude results in larger  excitation of $\rho_0$.  

\begin{figure}[t!]
	\begin{minipage}{3.5in}
		\includegraphics[scale=1.0]{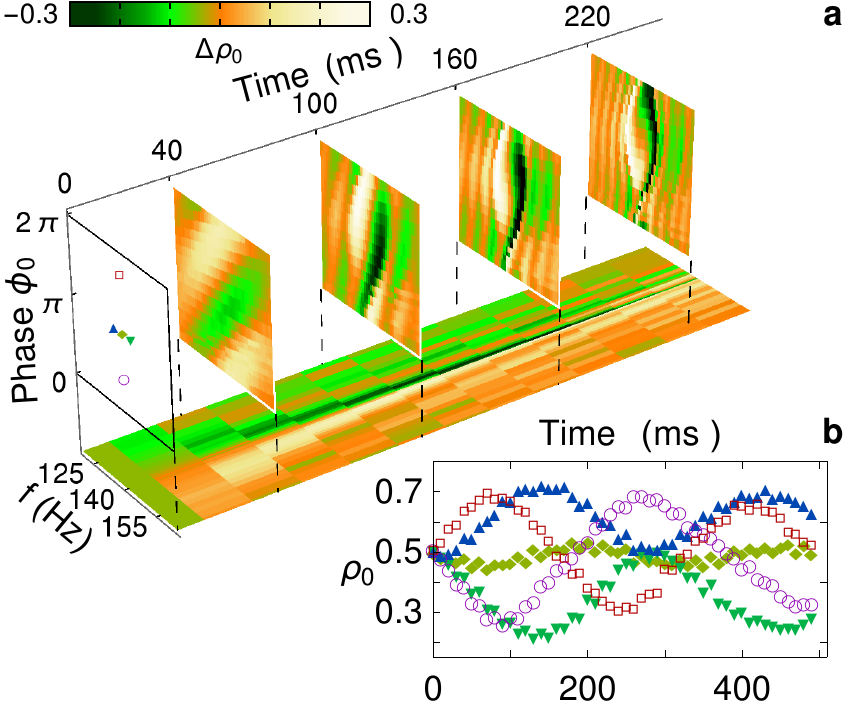}		
	\end{minipage}	
    \caption{\textbf{Excitation parameter map}. The data show the excitation of the condensate, $\Delta \rho_0 = \rho_0-\rho_0(0)$  following parametric excitation as a function of the excitation frequency $f$ and initial phase $\phi_0$ of the modulation. The initial state is $\rho_0(0),\theta_s = (0.5,\pi)$ and the modulation strength is $\epsilon= 0.5$
The four vertical slices show $\Delta \rho_0$ after 40~ms, 100~ms, 160~ms, and 220~ms of parametric excitation. The horizontal plot shows the evolution of $\Delta \rho_0$ for an initial phase $\phi_0=0$. The inset shows the temporal evolution of $\rho_0$ for different   $f,\phi_0$ pairs indicated in the vertical slice. The markers on the inset correspond to the markers on the main diagram.}
\label{fig:DelayModFreqMap}
\end{figure}

Parametric excitation is a coherent process, and hence it can be employed as a tool for  quantum control of the collective spin. However, accurate control requires detailed knowledge of the system response to the excitation parameters. In Fig. \ref{fig:DelayModFreqMap}, we present a parameter variation map that shows the excitation in the neighborhood of the $2f_0$ resonance for different values of excitation time $t$, phase $\phi_0$ and frequency $f$. 
For each measurement, we prepare the initial state $\rho_0,\theta_s=(0.5,\pi)$ using an rf pulse and modulate the quadratic Zeeman energy at different frequencies and initial phases. The change in population $\Delta\rho_0$ is measured after excitation times of 40~ms, 100~ms, 160~ms, and 220~ms as shown in Fig. \ref{fig:DelayModFreqMap} (four vertical slices). The white/orange (black/green) regions represent the positive (negative) changes in population. These two regions evolve and spiral to form a distinctive `yin-yang' pattern. This pattern in the measurements agrees well with theoretical calculations of the excitation (see Supplemental Information). At the center of the pattern $(f,\phi_0)=(2f_0,\pi)$ where $2f_0\approx 143$~Hz, $\rho_0$ remains unchanged and the excitation shows anti-symmetry about this point (diamond marker).  
The inset of Fig. \ref{fig:DelayModFreqMap} shows the temporal evolution of $\rho_0$ for different $f,\phi_0$ pairs indicated in the vertical slice. Although in each case the initial state is identical, the final state after excitation shows a strong dependence on both the frequency and phase of the applied modulation.  Additionally, we show a map of the population dynamics for the initial phase $\phi_0=0$ (Fig. \ref{fig:DelayModFreqMap}, horizontal slide). The two distinguishable domains, white (orange) black (green), are separated by the resonance frequency. The population dynamics exhibit oscillations during the excitation process, and, approaching the resonance frequency, both the oscillating period and the amplitude $(\Delta\rho_0)$ increase.

\begin{figure}[t]
	\begin{center}
	\begin{minipage}{3.5in}
		 \includegraphics[scale=0.9]{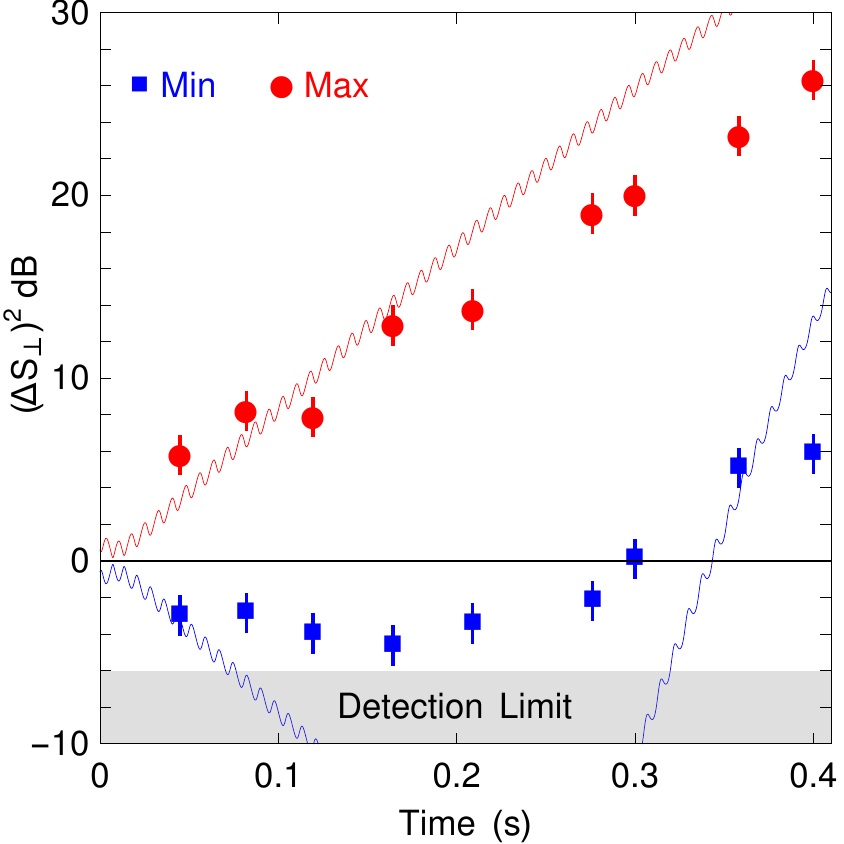}	
	\end{minipage}
	\end{center}
    \caption[Parametric excitation at the pole]{\textbf{Parametric excitation from $\rho_0=1$}.
Evolution of the minimum and maximum values of the transverse spin fluctuations, $\Delta S_\bot^2$, following parametric excitation with $\epsilon=0.56$ and $f=134$~Hz starting from the initial state $\rho_0=1$.  The measured maximum (red markers) and minimum (blue  markers) variance of  the transverse magnetization  $S_\bot$ is compared with simulation (solid red and solid blue).     
    }
\label{fig:SqueezingExcitationPole}
\end{figure}

We now turn to measurements of uniquely quantum features of the excitation. For a classical oscillator prepared in its stable equilibrium configuration, an important distinction between direct excitation and parametric excitation is that former can efficiently excite the oscillator while the latter cannot. The equilibrium (ground) state is a stable fixed point in phase space and if the oscillator is perfectly initialized it will remain unexcited by parametric modulation. However, for a quantum system prepared in its ground state, intrinsic Heisenberg-limited fluctuations of the state still allow for parametric excitation. In the semi-classical picture, the quantum fluctuations populate a family of orbits about the equilibrium point in the phase space that can be parametrically excited.

In Fig. \ref{fig:SqueezingExcitationPole}, we investigate parametric excitation from the quantum ground state of the condensate located at $\rho_0=1$ and demonstrate that parametric excitation can be used to generate quadrature squeezed states. 
Although the population, $\rho_0$, is largely insensitive to parametric excitation from the $\rho_0 = 1$ state (in contrast with the $\rho_0 \neq 1$ initial states), the fluctuations in the transverse coordinates, $ \Sperp_\bot ,  \Q_{\bot } $, evolve exponentially with time and show quadrature squeezing in the spin-nematic phase space. In contrast to our previous demonstration of squeezing \cite{HamleyNature2012}, in which the squeezing was generated by free dynamical evolution following a quench that localized the state at a unstable (hyperbolic) fixed point, here, the squeezing is generated near a stable fixed point by periodic distortion of the phase space orbits produced by the modulation of the quadratic Zeeman energy. 
The essential difference is the time-dependence in the Hamiltonian. 

In Fig. \ref{fig:SqueezingExcitationPole}, measurements of the minimum and maximum  values of the quadrature fluctuations of the transverse spin,  $\Delta S_\bot$, are shown and compared with a quantum simulation.
As evidenced by both the measurements and the calculations, the fluctuations in the initial state evolve exponentially at early evolution times and develop into quadrature squeezed states. The maximum squeezing measured is $-$5 dB, which is close to the detection-limited ceiling of $-$6 dB due to the photo-detection shot-noise and background scattered light \cite{HamleyNature2012}. The simulations suggest that with technical improvements, the system is capable of generating squeezing at the  $-$20 dB level. 
Although the experiments data show the main effects predicted by theory, the agreement of the measured fluctuations with the theory is not perfect, particularly at longer evolution times.  This is possibly due to effects of atom loss from the condensate, which has a lifetime of 1.5~s for these experiments, and we plan to further investigate this question in future work.

\noindent
\textbf{Discussion}

The parametric excitation can also be understood as transitions between eigentates of the many-body Hamiltonian, which can be calculated by diagonalizing the tridiagonal matrix  \cite{MiasPRA2008,GervingNature2012}
\begin{eqnarray}
	\label{BTriDiagonal}
	\mathcal{H}_{k,k'} &=& \{ 2 \tilde{c} k(2(N-2k)-1) + 2qk \} \delta_{k,k'} \nonumber \\
		 &+& 2 \tilde{c} \{ (k'+1) \sqrt{(N-2k')(N-2k'-1)} \delta_{k,k'+1} \nonumber \\
		 &+& k'\sqrt{(N-2k'+1)(N-2k'+2)} \delta_{k,k'-1} \}
\end{eqnarray}
written in the Fock basis $|N,M;k\rangle$ where  $k$ is the number of pairs of $m_f=\pm1$ atoms, $N$ is the total number of atoms and $M$ is the magnetization; both $N$ and $M$ are conserved by the Hamiltonian.  Treating $\tilde{c}$ as a perturbation, the eigenenergies are $E_k = 2qk+2\tilde{c} k(4N-4k-1)$ and the energy difference between Fock states is $\partial E_k/\partial k = 2q+2c(2\rho_0-1)$ (Supplemental Information).  Using this picture, we note that the parametric excitation spectrum excitation frequencies $f=2f_0/n$, $n=1,2,3...$ corresponds to many photon excitations of the system with  $f=2f_0$ being the single photon transition. 

It is interesting to contrast parametric excitation with the usual Rabi excitation of 2-level atomic spins using rf or microwave magnetic fields.
In parametric excitation, the time variation of a parameter modifies the Hamiltonian without displacing the equilibrium (ground) state of the system. As we have demonstrated, this can be achieved in a spin-1 condensate by simply modulating the magnitude of the bias magnetic field, which modulates the quadratic Zeeman energy term in the Hamiltonian. As the field strength is varied, the ground state remains at the pole of the sphere, while the shape of the orbits is modulated; this is similar to the case shown in \figref[a]{fig:Concept}. Rabi excitation of 2-level atomic spins using rf or microwave magnetic fields, on the other hand, is direct excitation rather than parametric excitation. Although in both cases the excitation employs time-varying magnetic fields, in the Rabi case the oscillating magnetic field is transverse to the bias field, which leads to a oscillation of the orientation of the total field. Because the ground state of the spin  aligns along the field direction, the addition of the time-varying transverse field leads to a periodic displacement of the ground state away from the pole in the usual Bloch sphere picture of the spin vector, similar to the case shown in \figref[b]{fig:Concept}.  

We point out that the techniques we have demonstrated are applicable to  related many-body systems including the double-well Bose-Hubbard (or Bosonic Josephon junction (BJJ) model), collisionally interacting pseudo-spin 1/2 two component condensates and ensembles of spin-1/2 atoms with photon mediated interactions.
Each of these systems can be described by a version of the Lipkin-Meshkov-Glick (LMG) model Hamiltonian \cite{lipkin65}, $H= U S^2_x - K S_z$, whose the mean-field phase space is functionally identical to the spin-nematic phase space shown in \figref[c]{fig:Concept} \cite{HamleyNature2012,VardiPRL2001}.
At their heart, these systems feature competing energy terms (one of which is non-linear) that give rise to a quantum critical point.  The $q > 2|c|$ polar phase that we explore in this work corresponds to the Rabi regime ($K > U$) in the BJJ system, which is the tunneling-dominated regime perturbed by the interactions $U S^2_x$, and the modulation of $q$ corresponds to a modulation of the tunneling coefficient $K$.
Indeed, there have been numerous theoretical proposals for excitations of these systems using periodic modulations (e.g. see \cite{Teichmann09,Boukobza10,Xie10,Jager15}), and many of the lattice-based experimental demonstrations mentioned previously realize closely related ideas generalized to multi-site systems \cite{StoferlePRL2004,ZenesiniPRL2009, SiasPRL2008,AlbertiNP2009,MaPRL2011,ChenPRL2011,HallerPRL2010,EckardtPRL2005}.

In summary, we have demonstrated a new mechanism for control and excitation of an ensemble of spins based on parametric excitation. This is a many-body control technique that relies on spin-dependent collisional interactions, which we have characterized for a wide range of control parameters.  We have shown that this method, when applied to the ground state, can be used to generate squeezed states.

\section{Methods}
\small

The experiment utilizes $^{87}$Rb atomic Bose-Einstein condensates created in an optical trap containing $N=40000$  atoms initialized in the $| F=1,m_F=0\rangle$ hyperfine state in a high magnetic field ($2$~G). To prepare the initial spin state, the condensate is rapidly quenched to a magnetic field of 1~G, and a Rabi rf pulse resonant with the $F=1$ Zeeman transition is applied to prepare the desired initial state $\rho_0,\theta_s=(\rho_0(t=0),\pi)$. The nominal value of the magnetic field $B=1$ G is determined by using rf and microwave spectroscopy, and the spinor dynamical rate is determined by measuring coherent spin dynamics oscillations using states prepared near the ferromagnetic ground state ($c=- 7(1)~\mathrm{Hz}$). 

Parametric excitation of the system is implemented by sinusoidally modulating $q$, which is implemented by time-varying the magnetic field using external coils.  Due to induced eddy currents in the metal vacuum chamber and the inductance of the magnetic field coils, the applied modulation is time-delayed relative to the intended control by 1~ms and reduced in amplitude by 15\%.  These effects are measured directly using magnetically sensitive rf spectroscopy of the atoms, and they are incorporated in all the simulations.   

The final spin populations of the condensate are measured  by releasing the trap and allowing the atoms to expand in a Stern-Gerlach magnetic field gradient to separate the $m_F$ spin components. The atoms are probed for $400~\mu\mathrm{s}$ with three pairs of counter-propagating orthogonal laser beams, and the fluorescence signal is collected by a CCD camera is used to determine the number of atoms in each spin component. 

To measure the transverse spin fluctuations $\Delta \Sperp_\bot$, an rf Rabi $\pi/2$ pulse is applied during the expansion  to rotate the transverse spin fluctuations (which are in the $x,y$ plane) into the $z$ measurement basis. The fluctuations are then determined from 30 repeated measurements of $\langle S_z \rangle$, the difference  in the number of atoms measured in the $m_F=1$ and $m_F=-1$ spin components.

{\bf Acknowledgements}  We acknowledge support from the National Science Foundation Grant PHY--1208828

{\bf Author Contributions}  T.M.H., C.D.H. and M.S.C. jointly conceived the study. T.M.H, M.A., B.A.R. and X.Y.Y. performed the experiment and analyzed the data. T.M.H., B.J.L. and C.D.H. developed essential theory and carried out the simulations. M.S.C supervised the work.

\normalsize\normalfont
\bibliography{PERefs}
\bibliographystyle{nature}

\onecolumngrid
\newpage
\section{Parametric Excitation of a Many-Body Spin System:  Supplementary Information}
\twocolumngrid
\begin{center}
\end{center}

In this Supplementary Information, we provide additional details on the theoretical calculations of parametric excitation in the spin-1 system using both  semi-classical and quantum approaches, and we present additional simulations to compare with the data shown in Fig. 3 of the main paper.
\section{Mean Field and Quantum Interpretation of Parametric Excitation}
\subsection{Coherent Oscillation Dynamics}
We first discuss parametric excitation using semi-classical mean field theory. The excitation occurs when the quadratic Zeeman energy is modulated at integer divisors of twice the natural coherent oscillation frequency in the $(\theta,\rho_0)$ phase space. The dynamics of the system are governed by a set of differential equations for the fractional population $\rho_0$ and the phase $\theta$ from \cite{ZhangPRA2005, ChangNature2005}
\begin{eqnarray}
	\label{Eqn:Meanfieldrhotheta}
	\dot{\rho}_0 &=&4\pi c\rho_0
	\sqrt{(1-\rho_0)^2-m^2}\sin 2\theta \\
	\dot{\theta} &=&-2\pi(q-c(1-2\rho_0))\\
	&\quad &-2\pi c\frac{(1-\rho_0)(1-2\rho_0)-m^2}{\sqrt{(1-\rho_0)^2-m^2}}\cos 2\theta. \notag
\end{eqnarray}
where $m=\langle \hat{S}_z \rangle$ and we have taken $h\rightarrow 1$. The spinor energy of the system is given by
\begin{eqnarray}
	\mathcal{E} &=& c\rho_0[ (1-\rho_0)+
	\sqrt{(1-\rho_0)^2-m^2}\cos 2\theta]
	+q(1-\rho_0)
	\notag 
\end{eqnarray}
and has an oscillation period \cite{ZhangPRA2005} of the form
\begin{equation}
	\label{Eqn:oscillationPeriod}
T=\frac{1}{\pi}\frac{\sqrt{2}}{\sqrt{-qc}}\frac{K(\sqrt{\frac{x_2-x_1}{x_3-x_1}})}{\sqrt{x_3-x_1}}
\end{equation}
where $K(k)$ is the elliptic integral of the first kind and $x_i$ are the roots of the differential equation
\begin{equation}
	(\dot{\rho}_0)^2 
= (4\pi)^2 ([\mathcal{E}-q(1-\rho_0)][(2c \rho_0+ q)(1-\rho_0)-\mathcal{E}]-(c\rho_0 m)^2). \notag
\end{equation}

For a condensate prepared in $m_F=0$, with magnetization conserved $m=0$, as is the case in our experiment, the roots $x_i$ are 
\begin{eqnarray}
	x_i & \in & \left\{
\frac{2c-q+\sqrt{4c^2-8c\mathcal{E}+4cq+q^2}}{4c}, ~  \frac{q-\mathcal{E}}{q},  \notag\right.\\
	&&\left.~\frac{2c-q - \sqrt{4c^2-8c\mathcal{E}+4cq+q^2}}{4c} \right\}. \notag
\end{eqnarray}

%

In order to calculate the period $T$, we first calculate
\begin{eqnarray}
	\frac{\sqrt{-qc}}{\sqrt{2}}\sqrt{x_3-x_1} 
	&=&\frac{\sqrt{q}}{2} \left(q^2+4qc(2\rho_0-1)+4c^2\right)^{1/4} \notag\\
	&\approx &\frac{\sqrt{q^2 +2qcQ_z}}{2}\left(1+\frac{ c^2(1-Q_z^2)}{(q+2cQ_z)^2}\right)
	\notag	
\end{eqnarray}
where $Q_z=2\rho_0-1$. The elliptical integral part of the period in Eq. \eqref{Eqn:oscillationPeriod}
\begin{eqnarray}
	K(\frac{x_2-x_1}{x_3-x_1}) \approx K(0.01) \approx \frac{\pi}{2}
\end{eqnarray}
Substituting these results back into the equation for the period, we obtain the natural  coherent oscillation frequency in quadrature phase ($\theta,\rho_0$)
\begin{eqnarray}
	f_0 &=& \frac{1}{T} =  \frac{\pi\sqrt{q^2 +2qcQ_z}}{2} \left(1+\frac{ c^2(1-Q_z^2)}
	{(q+2cQ_z)^2}\right) \frac{2}{\pi}\notag \\
	\label{Eqn:naturalFreq}
	&\approx &\sqrt{q^2 +2qcQ_z}
	\approx q+cQ_z
\end{eqnarray}
where we have that $q\sim 10|c|$ in our experiment.

Parametric excitation of the system is achieved by periodically modulating the quadratic Zeeman term $q$ in the Hamiltonian. Efficient excitation occurs when the modulation frequency is an integer divisor of twice the natural frequency of the system $f=\frac{2f_0}{n}$ where $n \in \mathbb{N}$. In our system, the coherent oscillations occur at a magnetic field $B=1$~G and spinor dynamical rate $c=-7.2(5)$~Hz. These parameters yield a range of natural frequencies
\begin{eqnarray}
	\label{Eqn:naturalFreqWithParams}
	f_0=71.6\times 1^2 -7.2 x \in [64.4,~78.8]~ \mathrm{Hz} ~~\forall \rho_0\in[0,~1].
\end{eqnarray}
with the most dominant excitation frequency corresponding to $n=1$
\begin{eqnarray}
	\label{Eqn:naturalFreqRange}
f=\frac{2f_0}{1}\in [128.8,~157.6]~~\mathrm{Hz}~\forall \rho_0 \in [0,~1].
\end{eqnarray}

\subsection{Parametric Excitation Theory}
In our experiment, the system is prepared in the $m_F=0$ state with magnetization $m=0$ and allowed to evolve at sufficiently high fields where $q/|c| = 10$. Under these conditions, Eq. \eqref{Eqn:Meanfieldrhotheta} simplifies to
\begin{eqnarray}
	\label{Eqn:Meanfieldrhotheta1G}
	\dot{\rho}_0 &=& 4\pi c\rho_0(1-\rho_0)\sin 2\theta \\
	\dot{\theta} &=&-2\pi(q 
	+  c(2\rho_0-1)). \notag
\end{eqnarray}
Parametric excitation is then applied by modulating the quadratic Zeeman energy
	\begin{equation}
		 q=q_0 +  \epsilon q_0 \sin (2\pi f_m t-\phi_0) H[t-\phi_0/2\pi f_m]. \notag
	\end{equation}
The Heaviside function and phase $\phi_0$ imply that the modulation starts at $q_0$ and is active for $t>\phi_0/2\pi f_m=\phi_0/\omega_m$, where $\omega_m=2\pi f_m$. We prepare an initial $\rho_0$ at high field using an rf pulse which initializes the quadrature phase $\theta_0=\pi/2$. The system then freely evolves for $t=\phi_0/\omega_m$, advancing the quadrature phase $\Delta\theta=-2\pi(q_0+2cx)\phi_0/\omega_m=-\omega_0\phi_0/\omega_m$, followed by modulation of $q$, where we have implicitly assumed that for evolution at high field we can set $\rho_0$ to be a constant so that integration of $\dot{\theta}$ is straightforward. We therefore have two initial contributions to the quadrature phase prior to modulation, namely $\theta_0$ and $\Delta \theta$.

For evolution times $t>\phi_0/\omega_m$ we can integrate $\dot{\theta}$ in Eq. \eqref{Eqn:Meanfieldrhotheta1G} giving:
\begin{eqnarray}
	\theta &=&\theta_0+\Delta\theta+\int_0^{t}\mathrm{d}t'\dot{\theta} \\
	&=&\pi/2-\omega_0\phi_0/\omega_m-\omega_0 t+\frac{2\pi\epsilon q_0}{\omega_m}\left(\cos(\omega_m t)-1\right)\notag
\end{eqnarray}
where $\omega_0=2\pi(q_0+c(2\rho_0-1))$. Substituting the phase into the population dynamical equation \ref{Eqn:Meanfieldrhotheta1G}, we obtain

\begin{eqnarray}
	\dot{\rho}_0 &=& 4\pi c\rho_0(1-\rho_0)\notag \\
	&&
	\times\sin \left[
	2\theta_0 -2\omega_0\phi_0/\omega_m-2\omega_0 t\right. \\
	&&\qquad +\left.\frac{4\pi\epsilon q_0}{\omega_m}\left(\cos(\omega_m t)-1\right)\right]\notag\\
	\label{Eqn:ParametricShapiro}
	&=& 4\pi c\rho_0(1-\rho_0) \sum_{n} J_n(\frac{4\pi \epsilon q_0}{\omega_m})\\
	&&\qquad \times\sin((n\omega_m-2\omega_0)t+\delta+n\pi/2) \notag 
\end{eqnarray}
where the phase $\delta=2\theta_0-2\omega_0\phi_0/\omega_m-4\pi\epsilon q_0/\omega_m$. In \eqnref{Eqn:ParametricShapiro}, we have made use of the identity $\sin(A+B)=\sin(A)\cos(B)+\cos(A)\sin(B)$ along with the Jacobi-Anger expansions
\begin{eqnarray}
	\cos(z \sin\alpha) = \sum_{n=-\infty}^\infty J_n(z) \cos(n\alpha) \notag \\
	\sin(z \sin\alpha) = \sum_{n=-\infty}^\infty J_n(z) \sin(n\alpha). \notag 		
\end{eqnarray}

Analyzing Eq. \ref{Eqn:ParametricShapiro} gives us some insight into the population dynamics as a function of the modulation parameters. When $\omega_m \neq 2\omega_0/n$, the time average of $\dot{\rho_0}\propto\sum_n\sin(\Omega t)$ is zero. When $\omega_m = 2\omega_0/n$, the time average of the $n$th term in the expansion $\dot{\rho}_{0,n} = 4\pi c\rho_0(1-\rho_0) J_n(\frac{4\pi \epsilon q_0}{\omega_m})\sin(\delta+\pi/2)$ is non-zero. Therefore, only for the case when $\omega_m = 2\omega_0/n$ is there sufficient coupling from the modulation to parametrically excite the system. The behavior of the Bessel functions $J_n\left(\frac{4\pi\epsilon q_0}{\omega_m}\right)$ also indicates that the strongest coupling occurs for $n=1$ or $\omega_m=2\omega_0$, a signature of parametric excitation.

The strength of the excitation is controlled by tuning $\epsilon$. When the system is modulated at $\omega_m=\omega_0/2n$ we can focus on short time dynamics $(\omega_m-2\omega_0)t<\pi$, since the higher order terms are negligible due to time averaging, and expand Eq. \ref{Eqn:ParametricShapiro} about $\epsilon=0$. It can be shown that the coefficient in the expansion is exact up to $\mathcal{O}(\epsilon)^{|n_{max}|}$. Using this fact, the population dynamical equation for $\dot{\rho_0}$ can be rewritten as:
\begin{eqnarray}
\label{Eqn:ParametricShapiroAltForm}
\dot{\rho_0}&=&4\pi c\rho_0(1-\rho_0)\\
&&\times\left[\sum_{n=0}^{\infty}\frac{(-1)^{n}x^{2n}}{(2n)!}\sin\delta+\sum_{n=0}^{\infty}\frac{(-1)^{n}x^{2n+1}}{(2n+1)!}\cos\delta\right] \notag \\ 
&&=4\pi c\rho_0(1-\rho_0)\left(\cos x\sin\delta+\sin x\cos\delta\right) \notag \\
&&=4\pi c\rho_0(1-\rho_0)\left(\sin\delta+x\cos\delta+\mathcal{O}(x)^2\right) \notag
\end{eqnarray}
where $x=4\pi\epsilon q_0/\omega_m$. Depending on the value of $\delta$, increasing the strength of the modulation $\epsilon$ will either enhance or reduce the response of the system to the modulation for short times. Furthermore, it is also apparent from Eq. \ref{Eqn:ParametricShapiroAltForm}, along with the expression for $\delta$, that the response of the system is periodic with respect to the initial phase of the modulation $\phi$, and has fixed points for $\rho_0=0.5$ and $\delta=n\pi$ where $n$ is an integer, both of which agree with the experimental data shown in Fig. 3 of the main paper.

We now consider the strong coupling case where $n=1$ or $\omega_m=2\omega_0$, and the higher order terms $|n|>1$ of $J_n(\frac{4\pi \epsilon q_0}{\omega_m})$ are negligible, so that
\begin{eqnarray}
	\label{Eqn:Rho0Freq}
	\dot{\rho}_0 &\approx & 4\pi c\rho_0(1-\rho_0) \notag\\
	&& \left( J_{-1}(\frac{4\pi \epsilon q_0}{2\omega_0})
 \sin(-4\omega_0 t+2\theta_0-\phi_0-\frac{4\pi \epsilon q_0}{2\omega_0}-\frac{\pi}{2}) \right. \notag \\
 &&+
	 J_{0}(\frac{4\pi \epsilon q_0}{2\omega_0})
 \sin(-2\omega_0 t+2\theta_0-\phi_0-\frac{4\pi \epsilon q_0}{2\omega_0}) \notag \\
 &&+ 
	\left. J_1(\frac{4\pi \epsilon q_0}{2\omega_0})
 \sin(2\theta_0-\phi_0-\frac{4\pi \epsilon q_0}{2\omega_0}+ \frac{\pi}{2}) \right) \notag 
\end{eqnarray}
and after integration 
\begin{eqnarray}
	\rho_0 &=& 4\pi c\rho_0(1-\rho_0) \notag\\
	&&\left(
	J_{-1}(\frac{4\pi \epsilon q_0}{2\omega_0}) \frac{1}{4\omega_0}
 \cos(-4\omega_0 t+2\theta_0-\phi_0-\frac{4\pi \epsilon q_0}{2\omega_0}- \frac{\pi}{2}) \right. \notag \\
 &&+
	 J_{0}(\frac{4\pi \epsilon q_0}{2\omega_0})
 \frac{1}{2\omega_0} \cos(-2\omega_0 t+2\theta_0-\phi_0-\frac{4\pi \epsilon q_0}{2\omega_0}) \notag \\
 &&+
	\left. J_1(\frac{4\pi \epsilon q_0}{2\omega_0})
 \sin(2\theta_0-\phi_0-\frac{4\pi \epsilon q_0}{2\omega_0}+ \frac{\pi}{2})t \right)+\rho_0(0).\notag 
\end{eqnarray}
The population $\rho_0(t) \approx \rho_0(0)$ if $\phi_0 = 1.37\pi$ and \mbox{$\omega_m=2\omega_0$}. 


\subsection{Fock States}
An alternative way to view the parametric excitation is by considering transitions between the eigenstates of the many-body Hamiltonian.  The energy corresponding to the oscillation frequency matches no single atom transition.  Rather, it approximately matches the energy difference between two atoms in the $m_F=0$ state and a pair of atoms in the $m_F=\pm1$ states.  Yet even this is not precise enough as this energy separation also depends on the collective state of the system varying from $\Delta\mathcal{E} \approx 2(q+c)$ for all atoms in the $m_F=0$ state to $\Delta\mathcal{E} \approx 2(q-c)$ for all atoms in the $m_F=\pm1$ states.  These energy separations can be calculated by diagonalizing the tridiagonal matrix given by
\begin{eqnarray}
	\label{BTriDiagonal}
	\mathcal{H}_{k,k'} &=& \{ 2 \tilde{c} k(2(N-2k)-1) + 2qk \} \delta_{k,k'} \nonumber \\
		 &+& 2 \tilde{c} \{ (k'+1) \sqrt{(N-2k')(N-2k'-1)} \delta_{k,k'+1} \nonumber \\
		 &+& k'\sqrt{(N-2k'+1)(N-2k'+2)} \delta_{k,k'-1} \}
\end{eqnarray}
where $\tilde{c}=c/2N$ and $k$ is the number of pairs of $m_F=\pm1$ atoms in the enumeration of the Fock basis.  The Fock basis, $|N,M,k\rangle$, is also enumerated with $N$ the total number of atoms, and $M$ the magnetization, both of which are conserved by the Hamiltonian leaving all dynamics in $k$. The off-diagonal contributions in Eq. \ref{BTriDiagonal} are due to the many body interaction given by the $\tilde{c} \hat{S}^2$ term of the Hamiltonian. This interaction results in mixing of the Fock states, even in the high field limit.  Without this interaction, there would be no transitions since the magnetic interactions, both linear and quadratic Zeeman, are diagonal in the Fock basis.  In this picture, the integer divisor frequencies of the spectrum correspond to many photon excitations of the system. 

\begin{figure}[t]
	\begin{center}
		\includegraphics[width=0.49\linewidth]{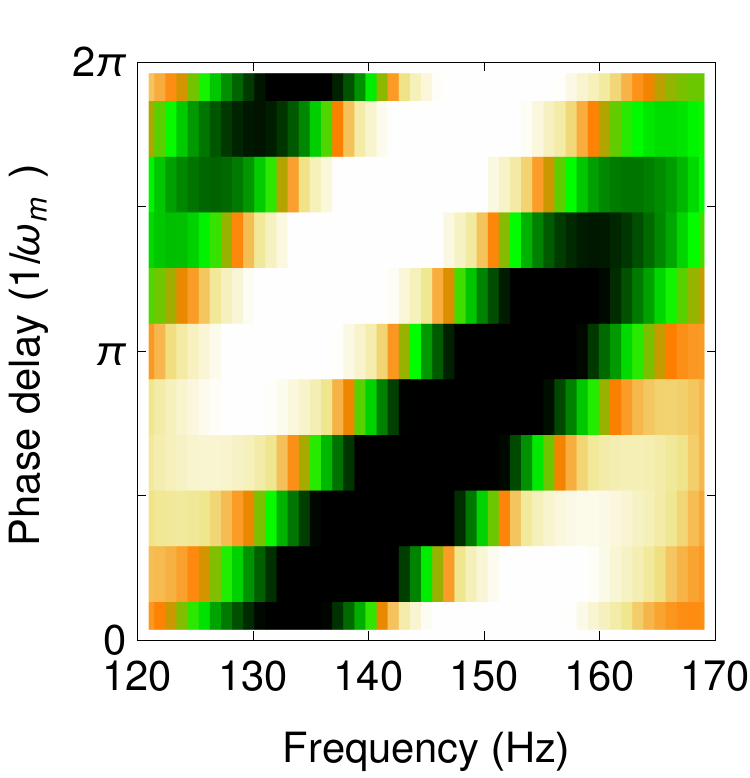}
		\includegraphics[width=0.49\linewidth]{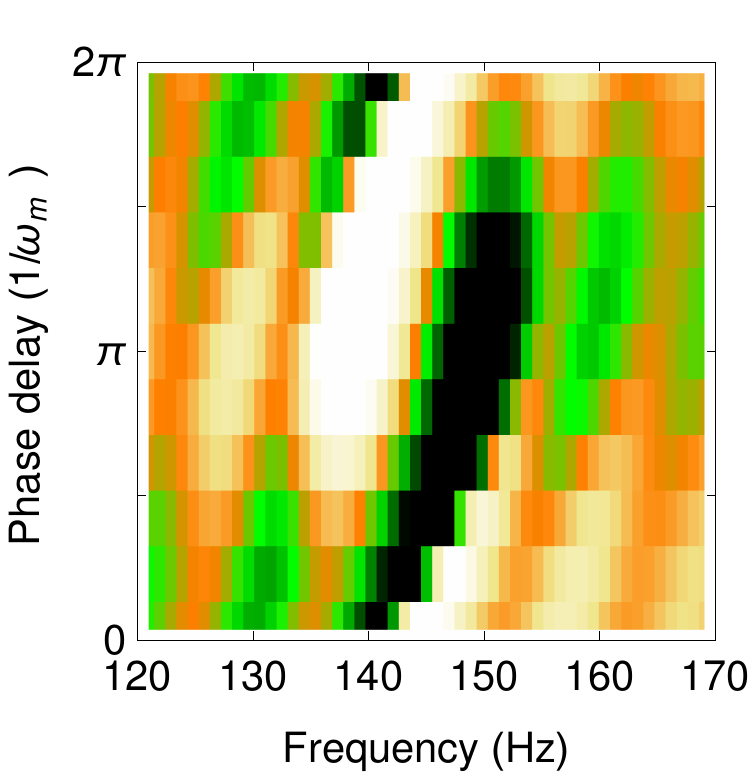} 	\\		
		\includegraphics[width=0.49\linewidth]{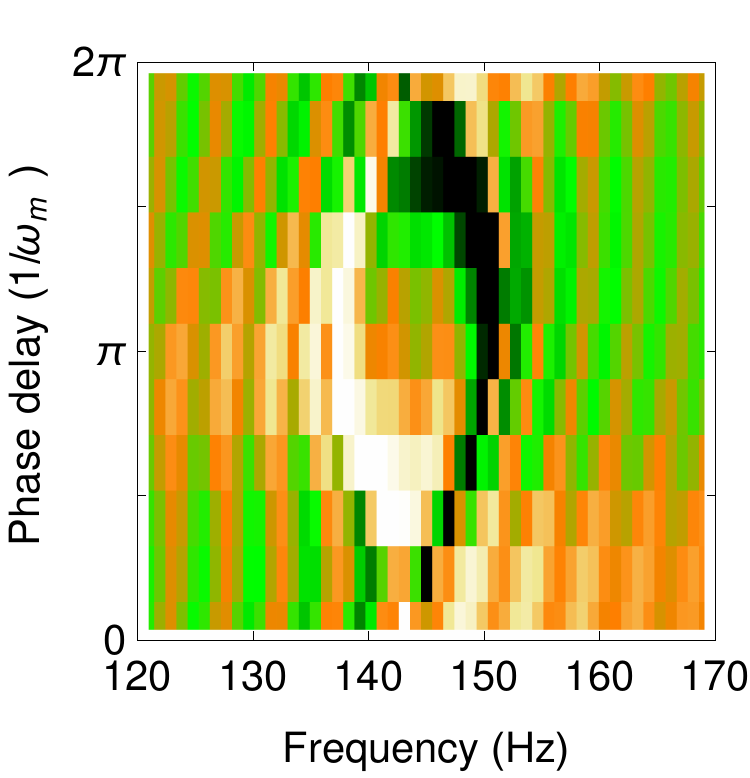} 	
		 \includegraphics[width=0.49\linewidth]{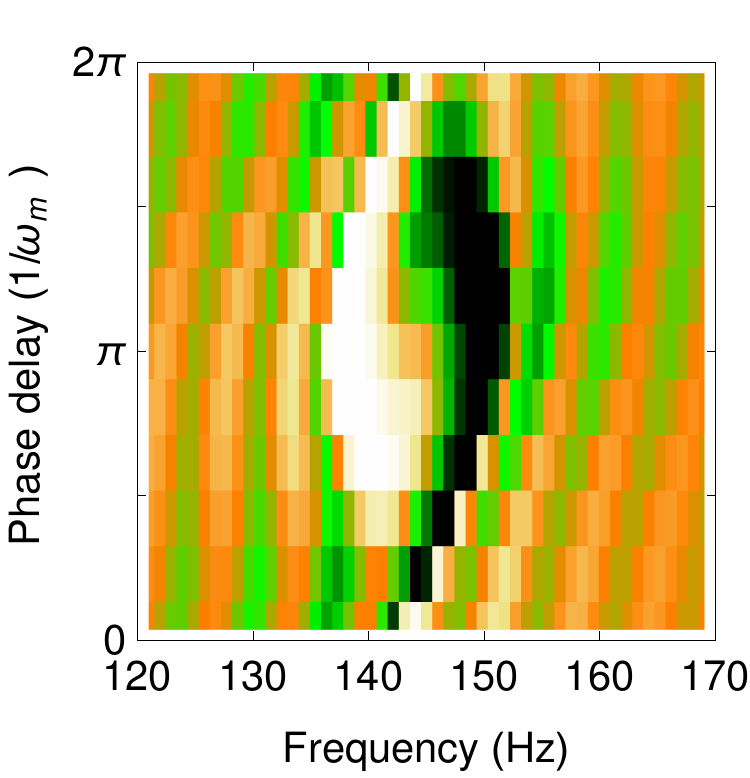} 	\\
	\end{center}
    \caption{\textbf{Excitation parameter map simulation}  Semi-classical simulations demonstrating parametric excitation for varying modulation phase $\phi_0$ and modulation frequency $f$ corresponding to evolution times of 40~ms, 100~ms, 160~ms, and 220~ms running clockwise. \label{fig:ParamVarMapSim}}
\end{figure}

In the regime where $q = q_Z B^2 > 2 c$, we treat $\tilde{c}$ as a perturbation and $\mathcal{H}^{(0)}_{k,k'} = 2 q k$ with expansion coefficients
\begin{eqnarray}
	E_k^{(0)} &=& 2 q k \notag\\
	E_k^{(1)} &=& \langle k \mid \mathcal{H'} \mid k \rangle = 2\tilde{c} k \left(2 (N - 2 k)-1\right) \notag
\end{eqnarray}
and eigenenergy of the system
\begin{eqnarray}
	E_k = E^{(0)}_k + E^{(1)}_k + \mathcal{O}^{(2)}.   \notag
\end{eqnarray}
The resonance frequency between Fock states is the energy difference between each Fock state
\begin{eqnarray}
	f &=& \frac{\partial E_k}{\partial k} 
	  = 2 q + 2\tilde{c}(2 N - 8 k -1) \notag \\
	  &\approx& 2(q+cx) \notag
\end{eqnarray}
where the last line corresponds to $k\ll N$. To first order, the resonance frequency between Fock states is the same as the frequency obtained from mean field theory Eq. \ref{Eqn:naturalFreq}. The factor of two arises from the definition of the resonant frequency $f=2f_0$.

\section{Parameter Variation Map Simulation}
In order to compare the experimental data shown in Fig. 3 of the main paper to theory, we perform four semi-classical simulations at fixed evolution times that demonstrate excitation in the neighborhood of the 2$f_0$ resonance for different values of the modulation phase $\phi_0$ and modulation frequency $f$. For each simulation, we initialize the state $(\rho_0,\theta_s)=(0.5,\pi)$ and modulate the quadratic Zeeman energy $q$ at different frequencies and phases. Details of the simulation method can be found in \cite{HamleyNature2012,GervingNature2012}.

The change in population $\Delta\rho_0$ is calculated after excitation times of 40~ms, 100~ms, 160~ms, and 220~ms, as shown running clockwise in Fig. \ref{fig:ParamVarMapSim}. These simulations correspond to the vertical slices shown in Fig. 3 of the main paper, and agree quite favorably. The white/orange (black/green) regions represent the positive (negative) changes in population which evolve and spiral to form a distinctive `yin-yang' pattern. At the center of the plots $(f,\phi_0)=(2f_0,\pi)$ where $2f_0\approx 143$~Hz, $\rho_0$ remains unchanged and the excitation shows anti-symmetry about this point.

\end{document}